\documentclass[prb,aps,twocolumn,showpacs, floatfix,preprintnumbers]{revtex4}

\usepackage[applemac]{inputenc}
\usepackage[T1]{fontenc} 
\usepackage{amsmath} 
\usepackage{subfigure}
\usepackage{lipsum}
\usepackage{amsfonts} 
\usepackage{amssymb, mathrsfs}
\usepackage{braket}
\usepackage{graphicx} 
\usepackage[usenames]{color} 

\usepackage{bbm}

\DeclareMathOperator\sn{sn}
\DeclareMathOperator\cn{cn}
\DeclareMathOperator\dn{dn}
\DeclareMathOperator\cd{cd}

\def\beq{\begin{equation}}
\def\eeq{\end{equation}}
\def\bsp{\begin{split}}
\def\esp{\end{split}}
\def\bea{\begin{eqnarray}}
\def\eea{\end{eqnarray}}
\def\ba{\begin{array}}
\def\ea{\end{array}}

\def\lb{\left(}
\def\rb{\right)}

\def\l.{\left.}
\def\r.{\right.}

\def\part{\partial}

\def\ket#1{\mid #1 {\cal{i}}}

\def\Z{\ensuremath{\mathbb{Z}}}

 \numberwithin{equation}{section}

\makeatletter

\newcommand{\Rmnum}[1]{\expandafter\@slowromancap\romannumeral #1@}
\makeatother
\begin{document}

\preprint{UdeM-GPP-TH-13-227}
\preprint{arXiv:1308.4395.}
\title{Quantum-classical phase transition of the escape rate of two-sublattice antiferromagnetic  large spins  }
\author{Solomon Akaraka Owerre}
\email{solomon.akaraka.owerre@umontreal.ca}
\author{M. B. Paranjape} 
\email{paranj@lps.umontreal.ca}
\affiliation{Groupe de physique des particules, D\'epartement de physique,
Universit\'e de Montr\'eal,
C.P. 6128, succ. centre-ville, Montr\'eal, 
Qu\'ebec, Canada, H3C 3J7 }

\begin{abstract}
\section*{Abstract}  
The Hamiltonian of a two-sublattice antiferromagnetic spins, with single (hard-axis) and double ion  anisotropies  described by $H=J \bold{\hat{S}}_{1}\cdot\bold{\hat{S}}_{2} - 2J_{z}\hat{S}_{1z}\hat{S}_{2z}+K(\hat{S}_{1z}^2 +\hat{S}_{2z}^2)$ is investigated using the method of effective potential. The problem is mapped to a single particle quantum-mechanical Hamiltonian in terms of the relative coordinate and reduced mass. We study the quantum-classical phase transition of the escape rate of this model. We show that the first-order phase transition for this model sets in at the critical value $J_c=\frac{K+J_z}{2}$ while for the anisotropic Heisenberg coupling $H = J(S_{1x}S_{2x} +S_{1y}S_{2y}) + J_zS_{1z}S_{2z} + K(S_{1z}^2+ S_{2z}^2)$ we obtain $J_c=\frac{2K-J_z}{3}$ . The phase diagrams of the transition are also studied.

\end{abstract}


\maketitle


\section{Introduction} 

The study of spin systems is a widespread subject ranging from the theory of magnetism, quantum spin hall effect, quantum computation, spintronics, superconductivity, to nuclear physics. In the last few decades, several methods have been developed to tackle the problem of spin systems. These methods include mapping to bosonic operators \cite{E,F,B} (e.g Schwinger Bosons, Holstein Primakoff transformation
 etc), semiclassical methods \cite{l,k,c,am,D} (e.g spin coherent state path integral) and effective potential method \cite{solo,solo1, wznw, solo2}. 
 
 In the last decade, there have been considerable interest in single ferromagnetic spin systems due to the fact that they exhibit first- or second-order phase transition between quantum and classical regimes for the escape rate \cite{solo4, cl,chud}. The transition occurs in the presence of a potential barrier and it takes place in two categories: Classical thermal activation over the barrier and quantum tunnelling through the barrier. The Classical thermal activation occurs at high temperature, in this case the transition rate is governed by $\Gamma \sim \exp\left[-\Delta V/T\right]$, where $\Delta V$ is the energy barrier. Below a particular temperature $T_{0}^{(0)}$, quantum tunnelling dominates thermal hopping and one should expect a temperature-independent rate of the form $\Gamma \sim\exp\left[-S_0\right]$, where $S_0$ is the Euclidean (imaginary time $t=-i\tau$) action evaluated along the instanton path. Equating the two exponents the crossover temperature (first-order transition) from quantum to classical regime is $T_{0}^{(0)} = \Delta V/S_0$. For a particle in a cubic or quartic parabolic potential interesting features arise, the so-called second order-phase transition at the temperature $T_{0}^{(2)}$. Below $T_{0}^{(2)}$  one has the phenomenon of thermally assisted tunnelling and above $T_{0}^{(2)}$ transition occur due to thermal activation to the top of the potential barrier\cite{chud, cl, solo4,chud1}. 

In the case of a uniaxial ferromagnetic spin model with a transverse magnetic field   Garanin and Chudnovsky\cite{cl} showed, by using the effective potential mapping \cite{solo,solo1, wznw}, that the phase transition can be understood in analogy of Landau's theory of phase transition, with the free energy expressed as $F = a\psi^2+b\psi^4 +c\psi^6$, where $a=0$ determines the quantum-classical transition and $b=0$ determines the boundary between the first- and second-order phase transition . The biaxial single ferromagnet spin has been studied by many authors \cite{solo2,solo4,solo6},   but not much is known about the behaviour of antiferomagnetic spin systems in this formalism. In this paper we consider a two-sublattice antiferromagnetic spins, denoted by $1$ and $2$, in the presence of a single (hard-axis) and double ion anisotropies. Such antiferromagnetic interactions arises in some compounds like CsFe$_8$, which was recently studied using the inelastic neutron scattering \cite{wal}.
The outline of this paper is as follows: In section II, we present the model Hamiltonian of the two-sublattice antiferromagnetic large spins, and then map it to a reduced one-dimensional quantum-mechanical particle using the well studied method of effective potential mapping. The instanton trajectory and the ground state energy splitting are also derived. In section III, we determine the condition for first-order phase transition in the spin model and finally some concluding remarks in section IV.

\section{Effective potential formalism}
The model we will consider in this paper is that of two-sublattice, antiferromagnetic, quantum spins in the presence of a single (hard-axis) and a double ion spin anisotropies. The corresponding  Hamiltonian is described by
\beq
H=J \bold{\hat{S}}_{1}\cdot\bold{\hat{S}}_{2} - 2J_{z}\hat{S}_{1z}\hat{S}_{2z}+K(\hat{S}_{1z}^2 +\hat{S}_{2z}^2)\label{1}
\eeq 
where $K, J_z $ $>0$ are the single and double ion anisotropy constants respectively, $J$ is the isotropic, antiferromagnetic, Heisenberg exchange constant and $\bold{\hat{S}}_{i}  (\text{ assumed to be large})$ $, i =1,2$ corresponds to the total spin on each sublattice.
 The spin operators obey the usual commutator relation: $\big[\hat{S}_{j\alpha},\hat{S}_{k\beta}\big]=i\epsilon_{\alpha \beta\gamma}\delta_{jk}\hat{S}_{k\gamma}$ $\left(j,k =1,2; \thinspace\alpha, \beta, \gamma  =x,y,z\right)$. The total spin commutes with the Hamiltonian only when $K=0=J_z$ but the total $z$-component of the spins $S_z = S_{1z}+ S_{2z}$ commutes with the full Hamiltonian.  Similar  models of this form has been extensively studied by using different methods\cite{los,ams,albert,had,wal}.
 
 
 Let us consider the problem of finding the spectrum of this present model \eqref{1}, the Hilbert space of the system describe by this Hamiltonian is the tensor product of the two spaces $\mathscr{H}_1 \otimes \mathscr{H}_2$.
In order to diagonalize the Hamiltonian  , let us first write its matrix representation in the basis of $S_{iz}$, given by  $\ket{s_1, m_1}\otimes\ket{s_2, m_2} =\ket{s_1s_2; m_1m_2}$. Introducing the eigenfunction 
 
\beq
\psi = \sum_{\substack {m_1=-s_1 \\ m_2=-s_2}}^{s_1,s_2} b_{m_1m_2} g_{m_1m_2}
\eeq
where 
\beq
g_{m_1m_2} = \binom{2s_1}{s_1+m_1}^{-1/2}\binom{2s_2}{s_2+m_2}^{-1/2}\ket{s_1s_2; m_1m_2}
\eeq
 we obtain
\begin{equation}
\begin{split}
H\psi=&\sum_{\substack {m_1=-s_1 \\ m_2=-s_2}}^{s_1,s_2}b_{m_1m_2}\Bigg[\frac{J(s_1-m_1)(s_2+m_2)}{2}g_{m_1+1,m_2-1}\\&+\frac{J(s_1+m_1)(s_2-m_2)}{2} g_{m_1-1,m_2+1}\\& +\Bigg((J-2J_z)m_1m_2+K(m_1^2 + m_2^2)\Bigg) g_{m_1m_2}\Bigg] = E\psi\end{split}
\label{6}
\end{equation}
which can be written equivalently as
\begin{equation}
\begin{split}
 E b_{m_1m_2}=& \left[(J-2J_z)m_1m_2+K(m_1^2 + m_2^2)\right]b_{m_1m_2}\\&+\frac{J(s_1-m_1+1)(s_2+m_2+1)}{2}b_{m_1-1,m_2+1}   \\&+\frac{J(s_1+m_1+1)(s_2-m_2+1)}{2} b_{m_1+1,m_2-1}  \end{split}
\label{6}
\end{equation}
where $b_{-s_i-1} = 0 = b_{s_i+1}$, etc, $i=1,2$. Introducing the generating function for the two particles \cite{solo1, wznw}
 
\begin{align}
\Phi(\phi_1,\phi_2) =\sum_{\substack {m_1=-s_1 \\ m_2=-s_2}}^{s_1,s_2} b_{m_1,m_2}e^{im_1\phi_1}e^{im_2\phi_2} 
  \label{7}
\end{align}
 
which obeys the periodic boundary condition 
\beq
\Phi(\phi_1+2\pi,\phi_2+2\pi) = e^{2\pi i (s_1+s_2)}\Phi(\phi_1,\phi_2)
\label{s1}
\eeq
The differential equation for $\Phi$ becomes

\begin{equation}
\begin{split}
&-K\left(\frac{d^2\Phi}{d\phi_1^2} +\frac{d^2\Phi}{d\phi_2^2}\right) -2 J \sin^2\left(\frac{\phi_1-\phi_2}{2}\right) \frac{d}{d\phi_1}\left(\frac{d\Phi}{d\phi_2}\right)\\& + 2J_z\frac{d}{d\phi_1}\left(\frac{d\Phi}{d\phi_2}\right)+ Js_1\sin(\phi_1-\phi_2)\frac{d\Phi}{d\phi_2}\\&-Js_2\sin(\phi_1-\phi_2)\frac{d\Phi}{d\phi_1} +Js_1s_2\cos(\phi_1-\phi_2)\Phi = E\Phi
\label{0} 
\end{split}
\end{equation}
Next we proceed in a similar fashion to that of two interacting classical particles by introducing the relative and the center of mass coordinates
\beq
r =\phi_1-\phi_2,\quad q = \frac{\phi_1 +\phi_2}{2}
\eeq
In this new coordinates, \eqref{0}  reduces to
\begin{equation}
\begin{split}
&G_1(r)\frac{d^2\Phi}{dr^2}  + G_2(r)\frac{d^2\Phi}{dq^2}+G_3(r)\frac{d\Phi}{dr}+ G_4(r)\frac{d\Phi}{dq} \\&+(G_5(r)- E)\Phi=0
\label{011} 
\end{split}
\end{equation}
The functions $G_i(r)$ are given by

\begin{equation}
\begin{split}
&G_1(r) =-2\left[K +J_z -J\sin^2\left(\frac{r}{2}\right)\right] \\&G_2(r) =-\frac{1}{2}\left[K +J_z+J\sin^2\left(\frac{r}{2}\right)\right] , \thinspace G_5(r)=Js_1s_2\cos r\thinspace 
\\& G_4(r)=\frac{J(s_1-s_2)}{2}\sin r, \thinspace G_3(r)=-J(s_1+s_2)\sin r\label{112}
\end{split}
\end{equation}
 where   $\Phi = \Phi(r,p)$. 
In order to make progress from \eqref{011}, we focus our attention on the case of equal spins i.e, $s_1=s_2$. 
Now since $G_4(r)$ vanishes, Eq.\eqref{011} can be simplified by separation of variable: $\Phi(r,p) = \mathcal{X}(r)\mathcal{Y}(q)$. The resulting equation becomes
  
\begin{equation}
\begin{split}
&\frac{1}{\mathcal{Y}(q)}\frac{d^2\mathcal{Y}(q)}{dq^2}+\frac{G_1(r)}{G_2(r)\mathcal{X}(r)}\frac{d^2\mathcal{X}(r)}{dr^2} +\frac{G_3(r)}{G_2(r)\mathcal{X}(r)}\frac{d\mathcal{X}(r)}{dr} \\&+\frac{(G_5(r)- E)}{G_2(r)} =0
\label{01} 
\end{split}
\end{equation}
Since the first term of the above expression is a function of $q$ only and  the rest of the terms are functions of $r$ only, both independent equations must be equal to a constant:
\begin{align}
&\frac{d^2\mathcal{Y}(q)}{dq^2} = \alpha\mathcal{Y}(q)\label{15}\\&
 G_1(r)\frac{d^2\mathcal{X}(r)}{dr^2} + G_3(r)\frac{d\mathcal{X}(r)}{dr} +{(G_5(r)- E)}\mathcal{X}(r) \label{16}\\&=\beta G_2(r) \mathcal{X}(r)
\nonumber
\end{align}
where $\alpha + \beta =0$.

There are three possible cases that satisfy this constraint:   $(i)$ Both $\alpha$ and $\beta$ are both zero, $(ii)$ $\alpha$ is a positive integer and $\beta$ is a negative integer, $(iii)$ $\alpha$ is a negative integer and $\beta$ is a positive integer. Due to the periodicity of the function $\Phi$, case $(ii)$ is not allowed. Cases $(i)$ and $(iii)$ are allowed since $\mathcal{Y}(q)= Aq +B$, with $A=0$ and $\mathcal{Y}(q)= A\cos(mq) +B\sin(mq)$, $m\in \Z$ obey the periodicity of $\Phi$. Using case $(iii)$, Eq.\eqref{16} can be written explicitly in the form
\begin{align}
&-2K(1+\delta)\left(1-\frac{\gamma}{1+\delta}\sin^2\left(\frac{r}{2}\right)\right)\frac{d^2\mathcal{X}(r)}{dr^2} - 2K \gamma s\sin r\frac{d\mathcal{X}(r)}{dr}\label{18}\\& \nonumber +K\Bigg(\gamma s^2 \cos r +\frac{m^2\gamma}{2}\sin^2\left(\frac{r}{2}\right)+\frac{m^2(1+\delta)}{2}\Bigg)  \mathcal{X}(r)\\&=E\mathcal{X}(r)\nonumber
\end{align}
The new parameters are defined as $\delta = J_z/K$, and $\gamma= J/K$.
The first derivative term can be removed by defining a new variable \cite{solo1,solo2}
\beq
 u =  F\left(\frac{r}{2} ,p\right)
=\int_0^r \frac{d r^{\prime}}{2\sqrt{1-p^2\sin^2\left(\frac{r^{\prime}}{2}\right) }} 
\label{jaco}
 \eeq
 which is the incomplete elliptic integral of the first kind with modulus $p^2 = \frac{\gamma}{1+\delta}$ and amplitude $\frac{r}{2}=$ am $u$ . The Jacobi elliptic sine and cosine are related to the trigonometric counterparts by $\sn u= \sin\left(\frac{r}{2}\right)$, $\cn u =\cos\left(\frac{r}{2}\right)$. We seek for a transformation of the form:
 \beq
 \mathcal{X}(r(u)) = \left(\dn u\right)^{(2s+ \frac{1}{2})} \Psi(u)
 \label{3.15}
  \eeq
 where $\dn u = \sqrt{1-p^2\sn^2u}$.
 
The function $\Psi(u)$ is regarded as the particle wavefunction since it tends to zero as $u\rightarrow \pm \infty$. One can show that  plugging \eqref{3.15} into \eqref{18} gives an equivalent  Schr\"{o}dinger equation:

 \beq
 H\Psi(u) = E\Psi(u)
\label{20}
\eeq
where 
\beq
H = -\frac{\nabla^2}{2\mu}  + V(u) , \quad \nabla =\frac{d}{du}
\label{3.18}
\eeq
The effective potential and the reduced mass are given by
\begin{align}
& V(u) = \frac{K \gamma}{8\dn^2 u}\Bigg[4\Bigg(\frac{m^2(1+\delta)}{\gamma}+2\left(s + \frac{1}{2}\right)^2\Bigg)\nonumber\\&+ \sn^2 u\Bigg((1+8  s (1 + s))\frac{\gamma}{1+\delta} -16\left(s + \frac{1}{2}\right)^2 \label{aka2} \\&+ \frac{\gamma-4m^2\gamma}{1+\delta} \sn^2u\Bigg)    \Bigg], \quad \mu  = \frac{1}{K(1+\delta)} \nonumber
\end{align}

Potentials of this form are complicated to deal with, however, further simplification can be made by using a well known approximation of the form\cite{wznw,solo1, kal}  $s(1+s) \sim \left(s +1/2\right)^{2} =\tilde{s}^2$. Hence, in a large spin system,  the terms independent of $\tilde{s}$ in the numerator of $V(u)$ make a very small contribution and thus can be ignored \cite{solo4}. The effective potential simplifies to

\begin{align}
V(u)=&K \tilde{s}^2\gamma +K \tilde{s}^2\gamma \frac{ 1+ \left(\frac{\gamma}{1+\delta}-2\right)\sn^2u  }{\dn^2u}\label{aka1}\\
&\nonumber=2K \tilde{s}^2\gamma\cd^2u; \quad \cd u = \frac{\cn u}{\dn u}
\end{align}
where a constant has been added to make the potential zero at the minimum. 
The effective potential is a periodic function with period $\pm2m\mathcal{K}\left(p\right)$,  and $\mathcal{K}\left(p\right)$ is the complete elliptic integral of first kind i.e $r=\pi$ in Eq.\eqref{jaco}. $u_{\text{min}}^{m}=\pm (2m +1)\mathcal{K}\left(p\right)$ corresponds to the minima while  $u_{\text{max}}^{m} = \pm 2m\mathcal{K}\left(p\right)$  corresponds to the position of the peaks. The height of the barrier given by
\beq
\Delta V=V_{\text{max}}-V_{\text{min}} = 2K \tilde{s}^2\gamma 
\eeq
The boundary condition for the particle wave function is
\beq
\Psi(u+2m\mathcal{K}\left(p\right)) =\Psi(u)
\eeq
This condition can be understood from \eqref{s1} as a consequence of $s_1=s_2$ which maps half-odd-integer spins to integer spins under a rotation of $2\pi$.


The Euclidean (imaginary time) Lagrangian corresponding to the Hamiltonian \eqref{3.18} is given by
\beq
L_{E} = \frac{1}{2}\mu  \dot{u}^2 + V(u)
\eeq
where $\dot{u}=\frac{du}{d\tau}$.
The Euler-Lagrange equation of motion is easily found as
\begin{align}
\mu\ddot{u} -V^{\prime}(u)=0
\end{align}
with
\beq
 \frac{1}{2}\mu \dot{u}^2 -V(u) =-E =0
\label{3.21}
\eeq
Integrating \eqref{3.21} we obtain the instanton
\begin{align}
&u(\tau) = \sn^{-1} \left[\tanh(\omega \tau)\right], \quad\omega^2= \frac{4K\gamma\tilde{s}^2}{\mu}
\label{aka5}
\end{align}
%

The instanton interpolates from the left minimum $u_i=u_{\text{min}}^{0} = -\mathcal{K}(p)$ at $\tau = -\infty$ to the center of the potential barrier $u_0=u_{\text{max}}^{0}=0$ at $\tau= 0$ and then arrives at the neighbouring right minimum $u_f=u_{\text{min}}^{0} = \mathcal{K}(p)$ at $\tau =\infty $. The corresponding instanton action is found to be 
\beq
S_0 = 2\tilde{s}\ln\left(\frac{1+ p}{1-p}\right),
\label{aka10}
 \eeq
where $p = \sqrt{\frac{\gamma}{ 1+\delta}}$.  
 
In order to understand the particle tunnelling in spin language. Consider Eq.\eqref{1} in terms of spherical coordinate (semi-classical approximation)\cite{c}. The potential energy corresponds to
\begin{align}
V &= Js^2\Bigg(\cos\theta_1\cos\theta_2 +\sin\theta_1\sin\theta_2\cos(\phi_1-\phi_2)\Bigg)\\& \nonumber-2J_zs^2\cos\theta_1\cos\theta_2+ Ks^2(\cos^2\theta_1+\cos^2\theta_2)
\end{align}
For $\theta_1=\pi/2=\theta_2$, the potential reduces to the form
\begin{align}
V &= 2K\gamma s^2\cos^2\Bigg(\frac{\phi_1-\phi_2}{2}\Bigg) \end{align}
 
  up to an additional constant. Thus, the ground state of the interacting system corresponds to $\phi_1-\phi_2=\pm \pi$. The full analysis of spin coherent state formalism is given in the appendix. The ground state tunnelling splitting can be computed with the help of \eqref{aka1}, \eqref{aka5} and \eqref{aka10} by 
using the standard procedure \cite{gag}.
This gives
\beq
\Delta E = \frac{2^{7/2}K(1+\delta)  (p\tilde{s})^{3/2}}{\sqrt{\pi (1 -p^2)}} \left(\frac{1- p}{1+p}\right)^{2\tilde{s}}
\eeq
%
\section{The condition for first order phase transition}
In this section we examine the possibility of first- and second-order phase transition in the two antiferromagnetic interacting spins under study. We will determine the condition at which first-order transition takes place and the temperature of the crossover for second order transition. At finite temperature the escape rate of a particle through a potential barrier in the quasiclassical approximation is\cite{chud, aff}
\begin{align}
\Gamma \sim \int dE \thinspace W(E) e^{-(E-E_{\text{min}})/T}
\end{align}
where $W(E)$ is the tunnelling probability of a particle at an energy $E$ and $E_{\text{min}}$ is bottom of the potential.  The tunnelling probability is defined via the imaginary time action \cite{sc}
\begin{align}
W(E) \sim e^{-S(E)}
\end{align}
Therefore we have
\begin{align}
\Gamma \sim e^{-F_{\text{min}}/T}
\end{align}
where $F_{\text{min}}$ is the minimum of the free energy $F \equiv E + TS(E)- E_{\text{min}}$ with respect to $E$. The imaginary time action is given by\cite{solo6,cl, chud1}
\begin{align}
S(E)= 2\int_{-u(E)}^{u(E)} dr\thinspace \sqrt{2m(V(u)-E)}
\label{4.4}
\end{align}
where $\pm u(E)$ are the turning points for the particles with energy $-E$ in the inverted potential $-V(u)$. The period of oscillation is given as $\tau(E)= -d S(E)/dE =1/T$. The first- and second-order transition now follows from the behaviour of the $\tau(E)$ as a function of $E$. Monotonically increasing $\tau(E)$ with the amplitude of oscillation gives a second-order transition while nonmonotonic behaviour of $\tau(E)$ ( that is a mininmum in the $\tau(E)$ vs $E$ curve ) gives a first-order transition\cite{solo4,cl,solo5,solo6, da}. For a constant mass, the  condition for the first-order phase transition can also be determined from the relation \cite{solo5}
\begin{align}
& -\frac{3\left(V^{\prime\prime\prime}(r_b)\right)^2}{8V^{\prime\prime}(r_b)}  +\frac{1}{8}V^{\prime\prime\prime\prime}(r_b) < 0
\label{3.17}
\end{align}
%

where $r_b$ corresponds to the top of the potential barrier (i.e the bottom of the inverted potential), in the present problem $r_b = u_{\text{max}}^{0}=0$
. Using \eqref{aka1} we obtain
the condition for first-order phase transition \eqref{3.17} as
\beq
 2K\gamma\tilde{s}^2\left[\left(1-\frac{\gamma}{1+\delta}\right)\left(1-\frac{2\gamma}{1+\delta}\right)\right] < 0
\eeq
It follows that the critical value is at $\gamma_c = \frac {1+\delta_c}{2}$.  Thus the first order phase transition occurs in the regime  $\frac {1+\delta}{2}<\gamma < 1+\delta$.
 Notice that the anisotropic Heisenberg coupling \beq H = \frac{J}{2}(S_1^{+}S_2^{-} +S_1^{-}S_2^{+}) + J_zS_{1z}S_{2z} + K(S_{1z}^2+ S_{2z}^2)\eeq
corresponds to the limit $\delta \rightarrow \frac{\gamma-\delta}{2} $. In this case the critical value becomes $\gamma_c = \frac{\left(2 - \delta\right)}{3}$. Thus, for $XY$$(\delta =0)$ coupling the critical value is $\gamma_c=\frac{2}{3}$, while for isotropic Heisenberg coupling $(\delta =\gamma)$ one obtains $\gamma_c=\frac{1}{2}$. In general, these critical values at which the first-order phase transition sets in can be written in terms of the modulus of the elliptic function $p^2$, with the critical value $p_{c}^2 = \frac{1}{2}$.
In the case of second-order transition the crossover occurs at $T_{0}^{(2)} = \omega_0/2\pi$, where $\omega_0= \sqrt{\lvert V^{\prime \prime}(0)\rvert/m}$ is the frequency of oscillation near the bottom of the inverted potential. Using the expressions in \eqref{aka2}  we obtain
\begin{align}
T_{0}^{(2)} = \frac{\omega_0}{2\pi} = \frac{\tilde{s}K(1+\delta)}{\pi}\sqrt{p^2\left(1-p^2\right)} 
\end{align}
The ground state crossover temperature below which quantum transition dominates is given by
\beq
T_{0}^{0} = \frac{V_{\text{max}}-V_{\text{min}}}{S_0}=\frac{\tilde{s}K(1+\delta)p^2}{\ln\left(\frac{1+p}{1-p}\right)}
\eeq

Introducing the following dimensionless parameters
\begin{align}
\vartheta =\frac{T}{T_{0}^{(2)}}, \quad Q = \frac{V_{\text{max}} - E}{\Delta V}
\end{align}
the effective free energy using \eqref{4.4} and \eqref{aka1} near the top of the barrier $(Q\ll1)$ takes the usual form \cite{cl}
\begin{align}
&F/\Delta V \cong
 1 + \left(\vartheta-1\right)Q +\frac{\vartheta}{4\left(1-p^2\right)}\left(\frac{1}{2}-p^2\right)Q^2 \\&
+ \frac{\vartheta}{8\left(1-p^2\right)^2}\left(p^4-p^2+\frac{3}{8} \right)Q^3 + O(Q^4)\nonumber
\end{align}
\begin{figure}[ht]
\centering
\subfigure[ ]{%
\includegraphics[scale=0.35]{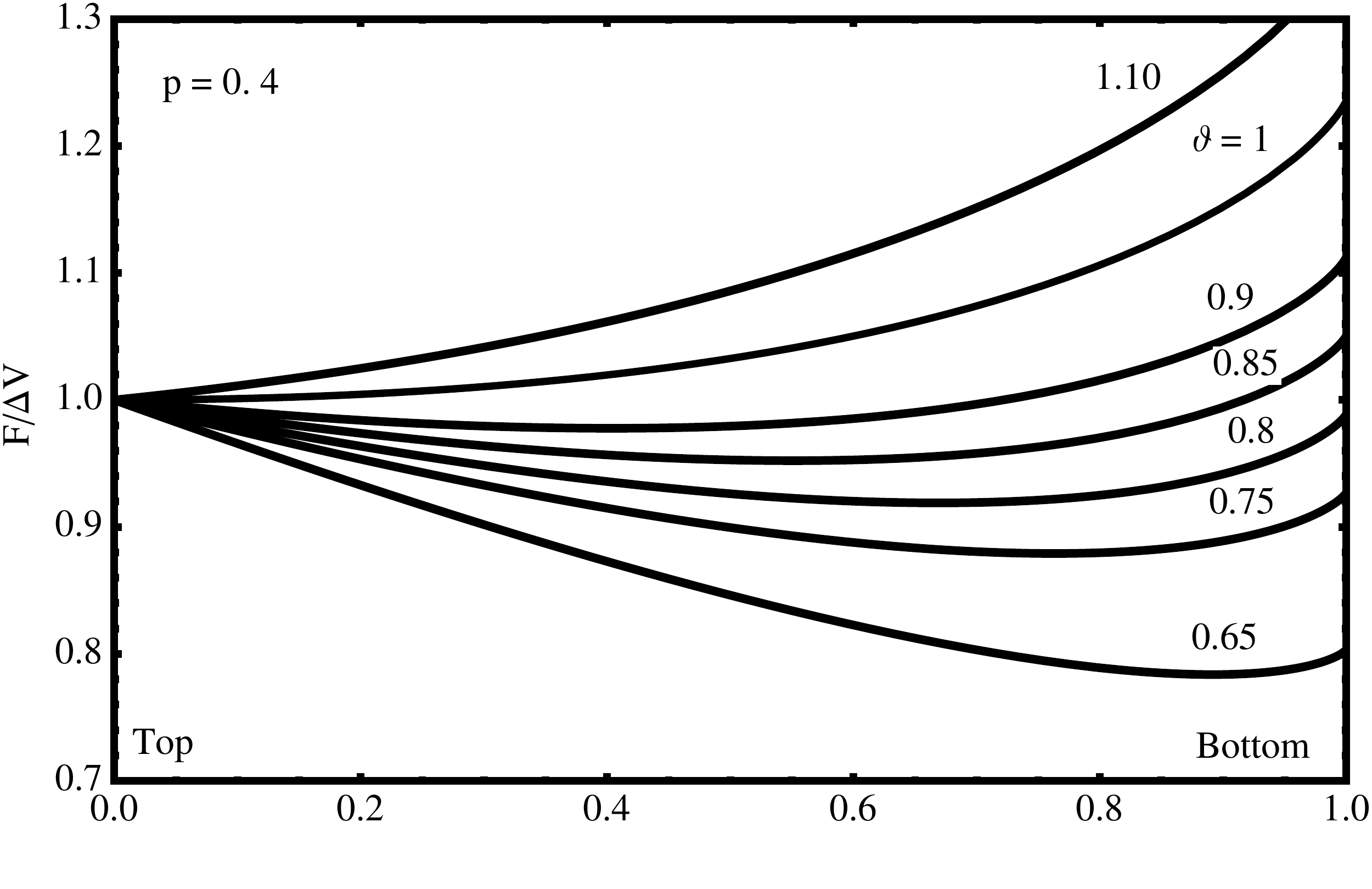}
\label{fr}}
\subfigure[]{%
\includegraphics[scale=0.35]{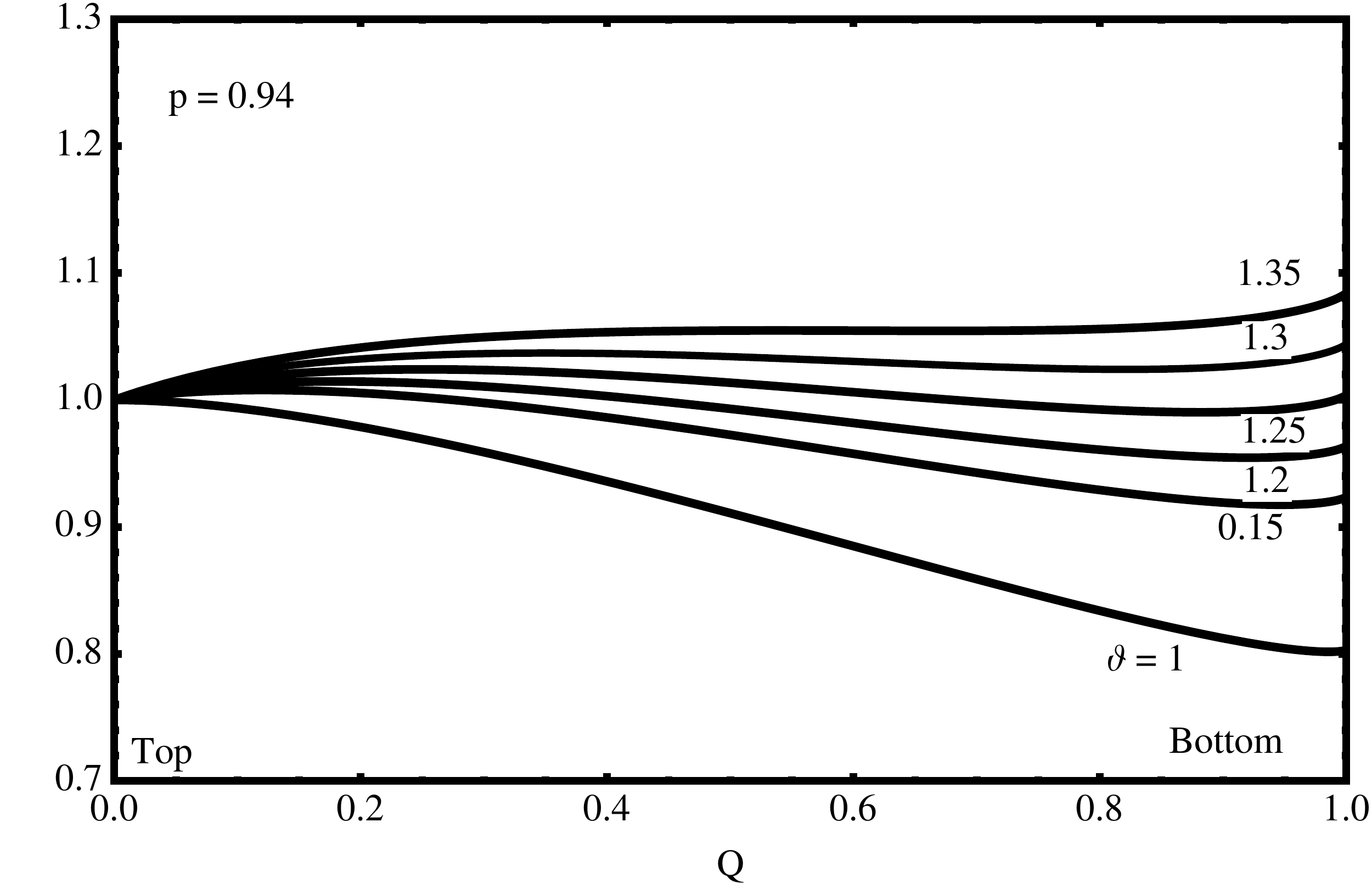}
\label{fr1}}
\caption{The effective free energy of the escape rate vs Q. (a) $p=0.4$, second-order transition, (b) $p=0.94$, first-order transition. }
\end{figure}
In analogy with Landau theory of phase transition, the boundary between the first-and second-order phase transition is again realized at $p_c^2=\frac{1}{2}$ corresponding to $\gamma_c = \frac {1+\delta_c}{2}$ as shown in Fig.\eqref{po}. The plot of the free energy against the parameter $Q$ for the whole range of energy is depicted in Fig.(1) for several values of $\vartheta$ at a particular $p^2$. It is shown that at $p = 0.4$, the minimum of $F$ does not change from $V_{\text{max}}-V_{\text{min}}$ for $\vartheta>1$. However for $\vartheta<1$ it drifts continuously from the top to the bottom of the potential which corresponds to the second-order transition from thermal activation to thermally assisted tunnelling (see Fig.\ref{fr}). At $p = 0.94$, $F$ possesses at least one minimum depending on the temperature. The crossover between classical and quantum regimes occurs when two minima have the same free energy. This is found to occur at $T_{0}^{(1)}= 1.25T_{0}^{(2)}$ (see Fig.\ref{fr1}).

\begin{figure}[h!]
\centering
 \includegraphics[scale=0.32]{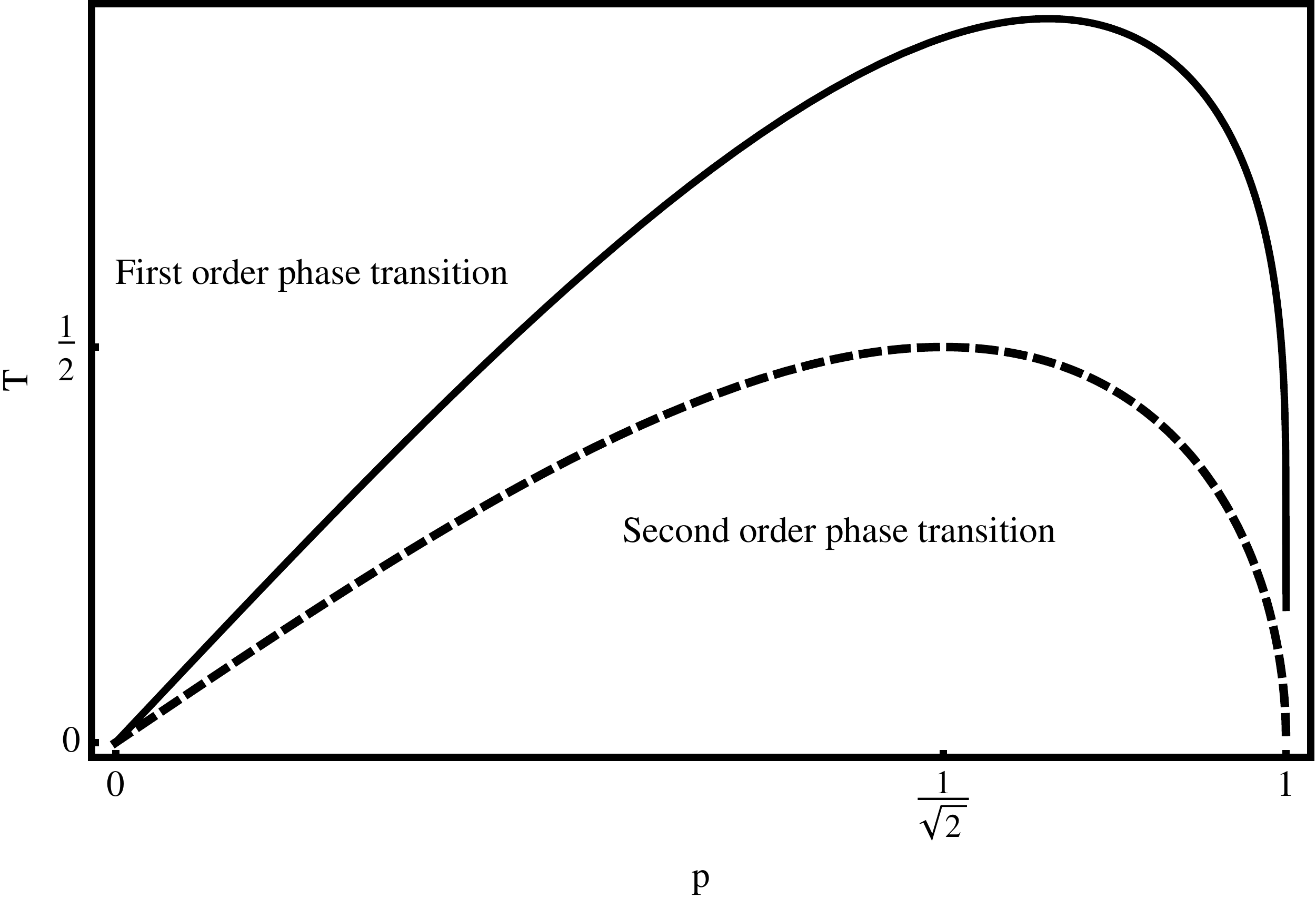}
\caption{The plot of the second order transition temperature (dashed line) and the cross over temperature (solid line) against $p$. Here $T = \pi/\tilde{s}K(1+\delta)T_0^{0,2}$}
\label{po}
\end{figure} 
\begin{figure}[ht]
\centering
\subfigure[ ]{%
\includegraphics[scale=0.3]{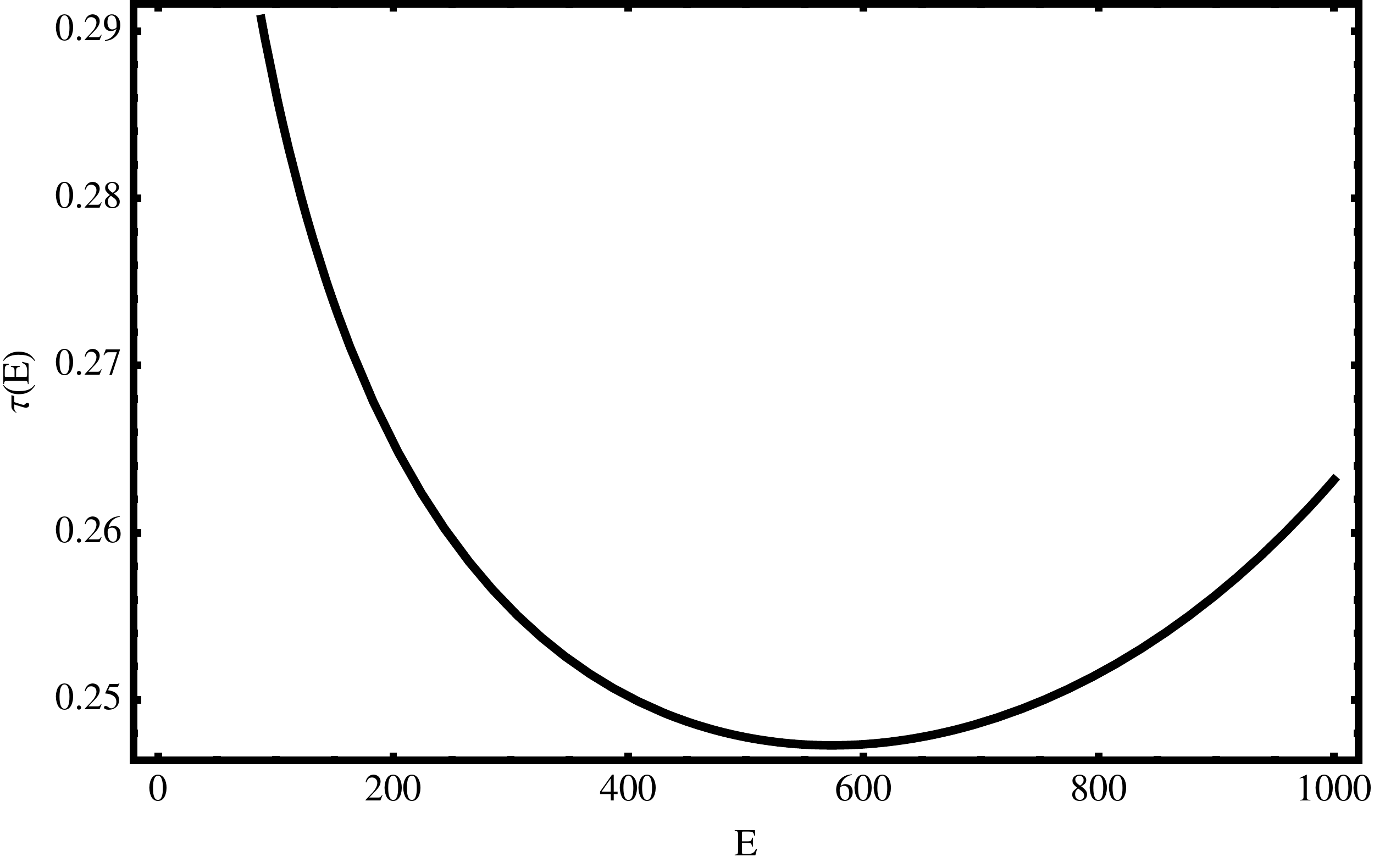}
\label{os}}
\subfigure[]{%
\includegraphics[scale=0.3]{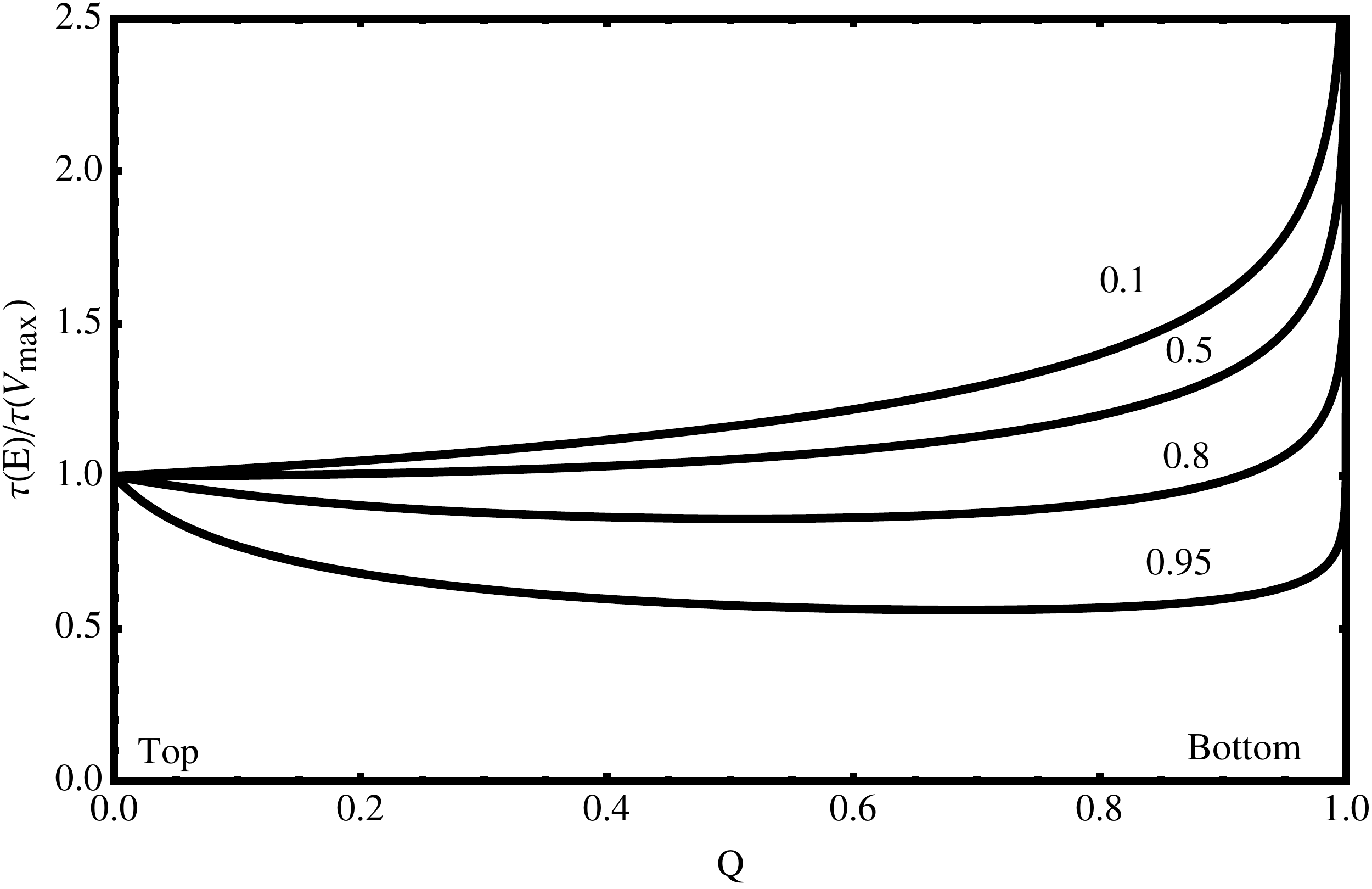}
\label{os1}}
\caption{(a): The plot of $\tau(E)$ against E   with $\gamma =0.85$ , $\delta =0.02$ , $K=1$ and $\tilde{s}^2=750$. (b): The plot of $\tau(E)/\tau(V_{\text{max}})$ against $Q$ for several values of  $p^2$ .  }
\end{figure} 

 The existence of first-order phase transition can also be seen in the $\tau(E)$ vs $E$ curve if the curve has a minimum and then rises again \cite{chud}. Using the method of periodic instanton (thermon) \cite{solo2,solo6,mula} that is by solving Eq.\eqref{3.21} with $E\neq0$, the period of oscillation is calculated to be
\begin{align}
&\tau(E) = \frac{4 \mathcal{K}(\kappa)}{\sqrt{2K(1+\delta)\left[2K\tilde{s}^2\gamma- p^2E\right]}}\label{5.15}\\& \kappa^2= \frac{b^2-1}{b^2-p^2},\quad b^2 = \frac{2K\tilde{s}^2\gamma}{E}
\nonumber
\end{align}

Near the top of the barrier $E\rightarrow 2K\tilde{s}^2\gamma$, $Q\rightarrow 0$ , $\kappa\rightarrow 0$ and $\mathcal{K}\rightarrow \pi/2$, the period reduces to $\tau = 2\pi/\omega_0$ while near the bottom of the barrier $E\rightarrow 0$, $Q\rightarrow 1$,
 $\kappa\rightarrow 1$ and the period diverges. The plot of the oscillation period against the energy is shown in Fig.\ref{os} for $\delta = 0.02$, $\gamma =0.85$, $K=1$, $p^2=0.83$ and $\tilde{s}^2=750$, the curve shows a minimum at $E_1\approx 550$, $\tau(E)\approx 0.24$ and then rises again at $E_0\approx 1000$, $\tau(E)\approx 0.26$ indicating that the first-order phase transition occurs for $\gamma > \frac{1+\delta}{2}$. Similar situation is depicted in Fig.\ref{os1} for $\tau(E)/\tau(V_{\text{max}})$ against $Q$, the nonmonotonic behaviour of $\tau(E)$ emerges above the  critical value $p^2_c=0.5$.
 \section{ Conclusions}
We have investigated the effective potential method of two-sublattice antiferromagnetic large  spins. The problem was mapped to a single particle Hamiltonian in terms of the relative coordinate and reduced mass. The instanton trajectory and the ground state energy splitting were found. It was found that the first-order phase transition of the escape rate kicked off at $\gamma_c=\frac{1+\delta}{2}$, $\frac{\left(2 - \delta\right)}{3}$ for isotropic and anisotropic interactions respectively, corresponding to $p^2_c=\frac{1}{2}$. At $p=0.4$, we found that the transition is of second-order with lowering temperature while at $p=0.94$ the transition is of first-order. The crossover between classical and quantum regimes occurred at $T_{0}^{(1)}= 1.25T_{0}^{(2)}$ for $p=0.94$. We hope that these  results can be experimentally investigated in some compounds that are described by the Hamiltonian \eqref{1} such as CsFe$_8$ etc.

\section{ Acknowledgments }
We thank  NSERC of Canada for financial support. 
\section{Appendix}
In the spin coherent state representation we have
\beq
S_E = \int d\tau \mathcal{L}_E
\eeq
The Euclidean Lagrangian is given by
\bea
& \mathcal{L}_E=is \dot{\phi_1}(1-\cos\theta_1)  +is \dot{\phi_2}(1-\cos\theta_2) \nonumber\\&+ Ks^2\Bigg[\gamma \left(\sin \theta_1\sin \theta_2 \cos(\phi_1-\phi_2)\right)\\&\nonumber-( 2\delta-\gamma)\cos \theta_1\cos \theta_2 + (\cos^2 \theta_1 +\cos^2 \theta_2)\Bigg]
\label{5}
\eea
The classical equations of motion for $\phi_1$ and $\phi_2$ are obtained by varying the action with respect to the variables. There are given by
\bea
 i\frac{d}{d\tau}\lb 1-\cos\theta_1\rb +Js\sin \theta_1 \sin \theta_2 \sin\lb \phi_1 -\phi_2 \rb =0
\label{7}
\eea
\bea
i\frac{d}{d\tau}\lb 1-\cos\theta_2 \rb -Js\sin \theta_1 \sin \theta_2 \sin\lb \phi_1 -\phi_2 \rb =0
\label{8}
\eea
For $\theta_1$ and $\theta_2$, we have the equations: 
\bea
 \frac{\delta S_E}{\delta  \theta_1} =0 =\frac{\delta  S_E}{\delta  \theta_2}
\label{9}
\eea
Adding  Eqn's \eqref{7} and  \eqref{8}  yields the conservation of total spin along the $z$ axis
\beq
 \cos\theta_1 +\cos\theta_2 = l=0 \quad
 \Longrightarrow \theta_1 = \pi -\theta_2
\label{11}
\eeq
where the constant $l$ is chosen to be zero using $ \theta_1 =\pi/2=\theta_2 $. Plugging Eq.\eqref{11} into the equations of motion and introducing the center of mass and relative coordinates used in the previous analysis: $\varphi= \phi_1-\phi_2$, and $\phi= \frac{\phi_1+\phi_2}{2}$. The resulting equations of motion can be derived from the Lagrangian

\beq 
\begin{split}
 \mathcal{L}_E=2is \dot{\phi} +is\dot{\varphi}\lb 1- \cos\theta\rb+V(\theta, \varphi)
\label{12}
\end{split}
\eeq
where 
\beq 
 V\lb\theta, \varphi\rb= 2K s^2 \Bigg[ \gamma\cos^2\frac{\varphi}{2}+ \left(1+\delta-\gamma \cos^2\frac{\varphi}{2}\right) \cos^2\theta\Bigg]
\label{13}
\eeq
A constant has been added to make the potential zero at the minimum $\theta_1=\theta = \pi/2$, $\varphi = \pm \pi$. The path integral becomes
\bea
\int\mathcal{D}\lb\phi\rb\mathcal{D}\lb\varphi\rb\mathcal{D}\lb\cos\theta\rb e^{-\int d\tau  \mathcal{L}_E}
\eea
Integrating out $\cos\theta$ we obtain
\bea
\int\mathcal{D}\lb\phi\rb\mathcal{D}\lb\varphi\rb  e^{-\int d\tau  \lb\mathcal{L}_{\text{cm}} +\mathcal{L}_{\text{r}}\rb}
\eea
where the center of mass and the relative coordinate Euclidean Lagrangians are given by

\begin{align}
\mathcal{L}_{\text{cm}} =2is\dot{\phi}, \quad\mathcal{L}_{\text{r}} = is\dot{\varphi}+\frac{\dot{\varphi}^2}{8K\lb 1 +\delta -\gamma\cos^2\frac{\varphi}{2}\rb} + 2K\gamma s^2\cos^2\frac{\varphi}{2}
\end{align}
The center of mass gives a phase in the path integral.  This can be related to the oscillation of spin wavefunction $\mathcal{Y}(q)$ in the previous analysis. The classical equation of motion for $\varphi$ gives
\bea
\dot{\varphi}^2 = 16K^2\gamma s^2\cos^2\frac{\varphi}{2}\lb 1 +\delta -\gamma\cos^2\frac{\varphi}{2}\rb
\eea
Integrating we obtain the instanton trajectory
\bea
\varphi(\tau) = 2\arcsin\Bigg[\frac{\lb 1+\delta -\lambda\rb^{1/2}\tanh\omega \tau}{\lb 1 +\delta-\lambda\tanh^2\omega \tau\rb^{1/2}}\Bigg]
\eea
The instanton interpolates from $\varphi=-\pi$ at $\tau=-\infty$ to $\varphi=\pi$ at $\tau=\infty$.
The corresponding action is
\bea
S_{\text{int}}= is \int_{-\pi}^{\pi}d\varphi +S_0
\eea
where
\begin{align}
S_0 &= s\sqrt{\gamma}\int_{-\pi}^{\pi} d\varphi \frac{\cos\frac{\varphi}{2}}{\sqrt{\lb 1 +\delta -\gamma\cos^2\frac{\varphi}{2}\rb}} = 2s\ln\lb\frac{1+\sqrt{\frac{\gamma}{1+\delta}}}{1-\sqrt{\frac{\gamma}{1+\delta}}}\rb
\end{align}
which is almost equal to \eqref{aka10} except for the quantum renormalization $\tilde{s}= (s+\frac{1}{2})$.

\end{document}